\documentclass[a4paper]{jpconf}

\usepackage{graphicx}
\begin{document}
\title{Looking for a charge asymmetry in cosmic rays}

\author{I Masina$^{1,2}$}

\address{$^1$ Dip.~di Fisica dell'Universit\`a di Ferrara and INFN Sez.~di Ferrara, Via Saragat, Ferrara, 
Italy} 
\address{$^2$ CP3-Origins, University of Southern Denmark,  Campusvej, Odense, Denmark}

\ead{masina@fe.infn.it}

\begin{abstract}
We combine the data from PAMELA and FERMI-LAT cosmic ray experiments by introducing a simple sum rule. 
This allows to investigate whether the lepton excess observed by these experiments is charge symmetric or not. 
We also show how the data can be used to predict the positron fraction at energies yet to be explored by 
the AMS-02 experiment. 
\end{abstract}

\section{Introduction}

The data collected by PAMELA \cite{Adriani:2008zr} indicate that there is a positron  
excess in the cosmic ray (CR) energy spectrum above $10$ GeV.
{The} rising behavior {observed by PAMELA} does not fit previous estimates of the CR formation 
and propagation implying the possible existence of a direct excess of CR {positrons} of unknown origins. 
Interestingly PAMELA's data show no excess in the anti-protons. 

{While ATIC \cite{:2008zzr} and PPB-BETS \cite{Yoshida:2008zzc} reported unexpected
structure in the all-electron spectrum in the range $100$~GeV- $1$~TeV, the
picture has changed with the higher-statistics measurements by
FERMI-LAT \cite{Abdo:2009zk} and HESS \cite{Aharonian:2008aa}, leading to a possible slight additional
unknown component in the CR $e^{\pm}$ flux over and above the specific Moskalenko and Strong model
prediction \cite{Strong:1998pw, Baltz:1998xv} which assumes a single-power-law injection spectrum.}

These interesting features have drawn much attention, and many explanations have been proposed: {} For example,
these excesses could be due to an inadequate account of the CR background 
in previous modeling; The presence of new astrophysical sources;
They could also originate from annihilations and/or decays of dark matter. 
We refer to \cite{Fan:2010yq} for a recent review. 

Whatever the origin of these excesses might be, following Ref. \cite{Frandsen:2010mr}, 
we show how to derive constraints about their charge asymmetry. Clearly, this will
eventually shed some light on the physical nature of their source.

\section{A useful sum rule}

We start by writing the observed flux of electrons and positrons as the sum of two contributions:
A background component, $\phi_{\pm}^B$, due to known astrophysical sources (which actually 
are better known for the electrons rather than the positrons), 
and an unknown component, $\phi_{\pm}^U$, in formulae:
\begin{equation}
\phi_\pm = \phi_\pm^U + \phi_\pm^B  \ .
\end{equation}
The component $\phi_{\pm}^U$ is the one needed to explain the features in the spectra 
observed by PAMELA and FERMI-LAT. 

These experiments measure respectively the positron fraction and the total electron and positron fluxes 
as a function of the energy $E$ of the detected $e^\pm$, i.e.: 
\begin{equation}
P(E) = \frac{\phi_+(E)}{\phi_+(E) + \phi_-(E)}\ , \qquad  F(E) = \phi_+(E) + \phi_-(E) \ .
\end{equation}
The left-hand side of the equations above refer to the experimental measures. 
The contribution from the unknown source is then: 
\begin{eqnarray}
\phi_+^U(E) & = &  P(E)~F(E) -\phi_+^B(E) \ ,  \\ 
\phi_-^U(E) & = & F(E)~ \left(1-P(E)\right) -\phi_-^B(E) \ .
\end{eqnarray}
In terms of their difference and sum:
\begin{eqnarray}
\phi_+^U(E) -\phi_-^U(E) &=&   F(E)~(2 P(E)-1) +(\phi_-^B(E) -\phi_+^B(E) ) \ ,\nonumber \\  
\phi_+^U(E) +\phi_-^U(E) &=&  F(E)- (\phi_-^B(E)+\phi_+^B(E)) \nonumber \ . \\
&& 
\label{sumrules}
\end{eqnarray}
The latter equation implies $F(E)\geq \phi_-^B(E)+\phi_+^B(E)$.
We model the background spectrum using $\phi_\pm^B(E)=N_B B^\pm(E)$, where
$N_{B}$ is a normalization coefficient such that $F(E)/ (B^-(E)+ B^+(E))\geq N_B$
and $B^\pm(E)$ are provided using specific astrophysical models. 
In this paper we adopt the popular Moskalenko and Strong model \cite{Strong:1998pw, Baltz:1998xv},
for which $N_B$ is less than  $0.75$  and $B^\pm(E)$ are given by: 
\begin{eqnarray}
B^+&=&  \frac{ 4.5 E^{0.7}}{1 + 650E^{2.3} + 1500E^{4.2}} \ ,
 \\
B^-&=& B_1^- + B_2^- \ ,\\ 
B_1^-&=& \frac{ 0.16 E^{-1.1}} {1 + 11 E^{0.9} + 3.2 E^{2.15}}  \ , 
  \\
B_2^-&=& \frac{ 0.70 E^{0.7}}{1 + 110 E^{1.5} + 600 E^{2.9} + 580 E^{4.2}} \ ,
\end{eqnarray}
where $E$ is measured in ${\rm GeV}$ and the $B$s in ${\rm GeV}^{-1} {\rm cm}^{-2}{\rm sec}^{-1}{\rm sr}^{-1}$ units. We checked that our results remain unchanged when replacing the parameterization above with the one adopted by the Fermi Collaboration (model zero) \cite{Grasso:2009ma, Ibarra:2009dr}. 

It is convenient to introduce the following parameter: 
\begin{equation}
r_U(E) \equiv \frac{\phi_-^U(E)}{\phi_+^U(E)}= \frac{F(E) ~(1-P(E))-\phi_-^B(E)}{P(E)~F(E)-\phi_+^B(E)}~~,
\label{ruu}
\end{equation}
which quantifies the level of charge asymmetry in the unknown component of electrons and positrons.
The equation above can be rewritten as 
\begin{equation}
R(E) \equiv \frac{F(E)}{B^-(E)} ~\frac{ 1-(1+r_U(E)) P(E)}{1-r_U(E) \frac{\phi_+^B(E)}{\phi_-^B(E)}} = N_B \ .
\label{sumrulefinal}
\end{equation}
Although the sum rule $R(E)$ seems to depend on the energy it should, in fact, be a constant 
as is clear from the right hand side of the previous equation.  
{\it This leads to a nontrivial constraint linking together in an explicit form the experimental results, 
the model of the backgrounds and the dependence on the energy of the unknown components.}  

We now turn to the actual data and show in which way the sum rule
provides essential information on the unknown components of the CRs.

Since we use simultaneously the results of FERMI-LAT and PAMELA we are obliged to consider only the common 
energy range.  Note that the CRs energy range below $10-20~{\rm GeVs}$, where the spectrum 
is affected by the Sun, is outside the common range. 
Within the relevant  but limited range of energies we will consider here it is therefore sensible 
to assume $r_U$ to be nearly constant. 

\begin{figure}[h!]\vspace*{.5cm} 
\begin{center} 
\includegraphics[width=7.cm]{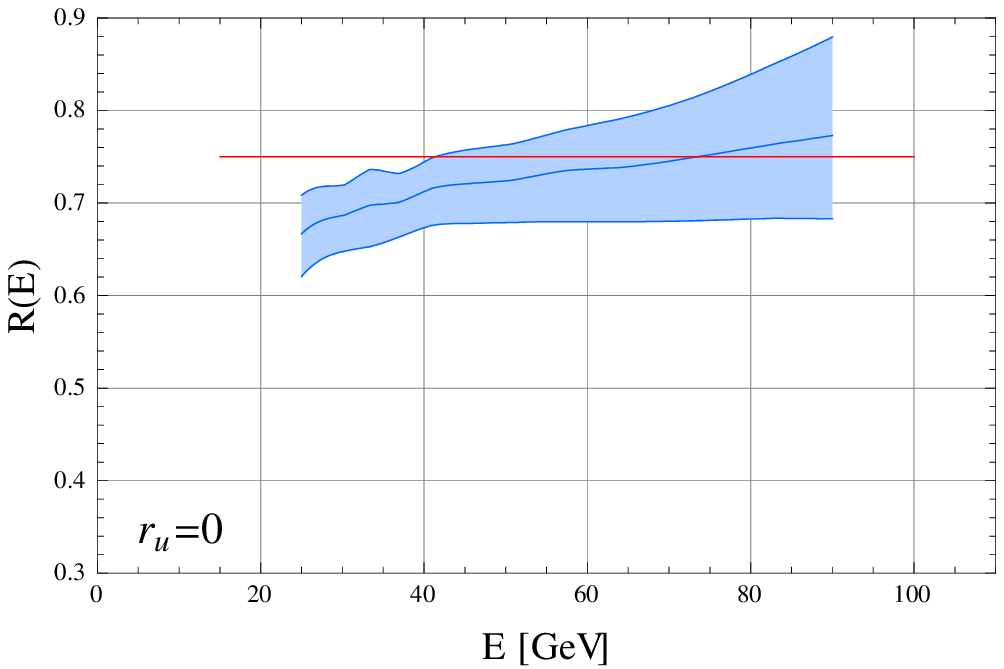}~~~ 
\includegraphics[width=7.cm]{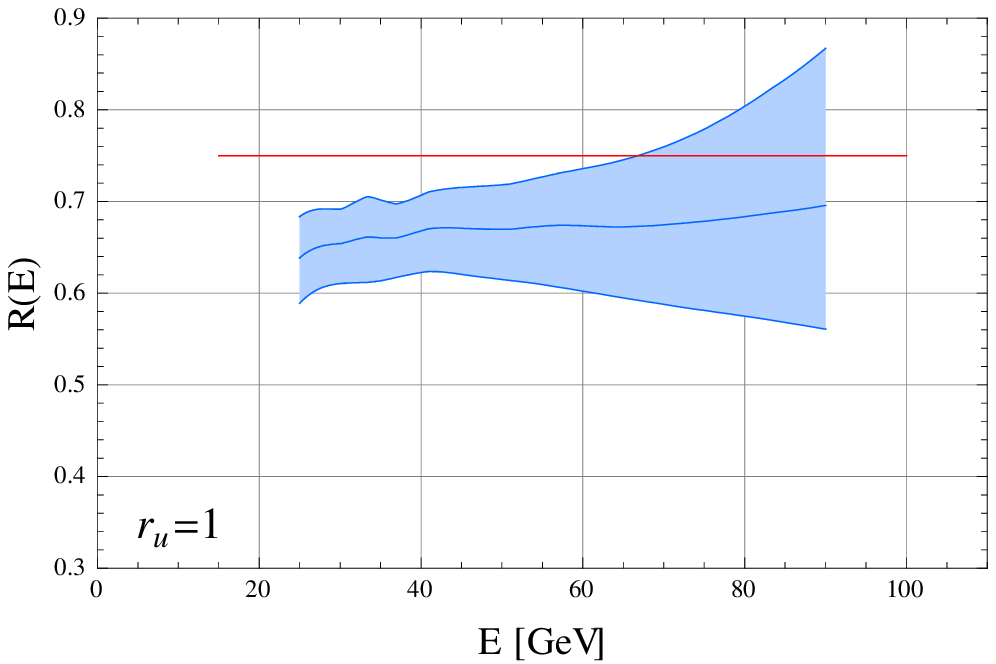} \\
\includegraphics[width=7.cm]{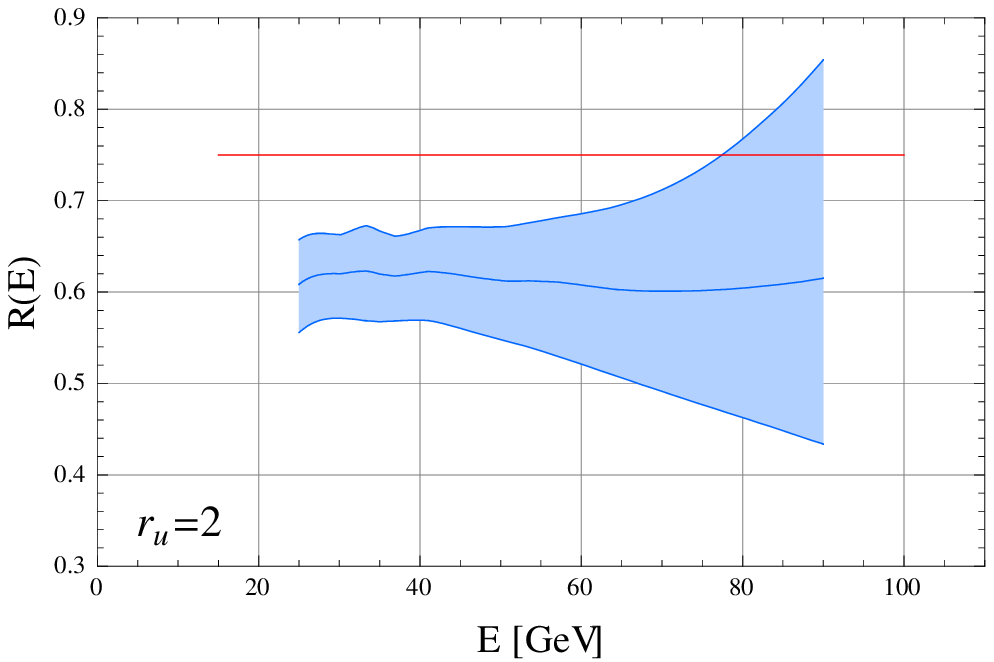} ~~~
\includegraphics[width=7.cm]{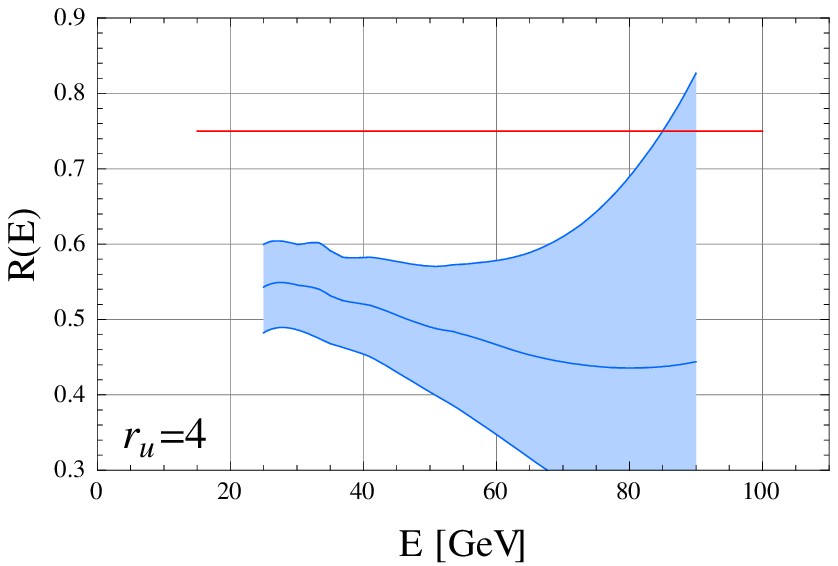} 
\end{center}
\vspace*{0.5cm} 
\caption{\label{Ratios} Ratio $R(E)$ as a function of the energy $E$ of electrons and positrons
and for values of $r_U=0,1,2,4$.
The shaded region accounts for the $1\sigma$ error in PAMELA and FERMI-LAT data (2010 release). 
Secondaries are estimated according to the expressions in \cite{Strong:1998pw, Baltz:1998xv}.
The value of $R(E)$ thus corresponds to the normalization factor for the background, called $N_B$ 
in the text.}
\end{figure}

We find useful to plot the function $R(E)$ for different values of $r_U$  in order to test the 
sum rule. This would imply that this function is independent of the energy. 
The associated constant value would then be identified with the background CRs normalization factor $N_B$. 
We report the results in Fig.~\ref{Ratios}. The straight (red) line is the $N_B = 0.75$ value which is 
the largest one can assume for the background not to be larger than the FERMI-LAT results. 
The shaded region is obtained by using the PAMELA and FERMI-LAT data released in 2010, 
assuming a $1\sigma$ error band for both.
We observe that there is a clear tendency for the combined data to predict a lower value of 
the constant $N_B$, say about $0.66, 0.64, 0.62, 0.55$ for increasing value of the ratio $r_U = 0,1,2,4$. 
This is clear when looking, from top to bottom, at the different panels of Fig.~\ref{Ratios}. 
It is interesting to note that we find a plateau, in the relevant energy range, up to $r_U$ near 
the value of $4$ when $R(E)$ starts showing some deviation.  

Given the large uncertainties (for PAMELA, the 2010 $1\sigma$ error bands are larger than in 2008) we cannot 
yet provide a more solid conclusion, however we can use the derived normalization $N_B$ for each different 
ratios of the unknown components to {\it predict} the positron fraction at energies higher than 
the ones provided so far by PAMELA.

\section{PAMELA and the charge asymmetry of the unknown component}

\begin{figure}[h!]\vspace*{0.5cm} 
\begin{center} 
\includegraphics[width=12cm]{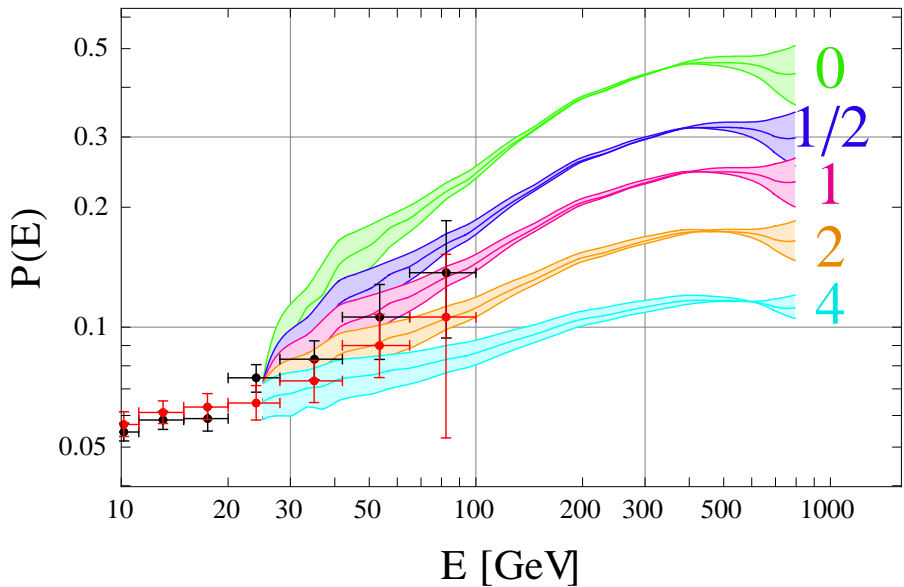} 
\end{center}
\caption{Model independent prediction for the positron fraction $P(E)$ as a function of the energy $E$ 
of electrons and positrons and for vales of $r_U=0,1/2,1,2,4$, from top to bottom.
Secondaries are estimated according to the expressions in \cite{Strong:1998pw, Baltz:1998xv} reported in 
the main text. The shaded regions account for the range of the allowed $N_B$ values (see Fig. 1),
obtained in turn by considering the 2010 PAMELA and FERMI-LAT data at $1\sigma$.
We also display the 2010 (red) and 2008 (black) PAMELA data, with $1\sigma$ error bars.}
\label{predictionP}
\end{figure}

In order to be able to make such a prediction we first rewrite eq. (\ref{sumrulefinal}) as follows:
\begin{equation}
P(E) = \frac{1}{1+r_U} \left(1-\frac{\phi^B_-(E)}{F(E)}(1-r_U \frac{\phi^B_+(E)}{\phi^B_-(E)})\right) \ , 
\end{equation}
where we use for each $r_U$ the estimated associated $N_B$ (referring to Fig. 1, we consider 
the range of $N_B$ values allowed for $E\approx 25$ GeV). The different predictions for the positron 
fraction, assuming that $r_U$ remains constant over the entire energy range, 
up to $1000$~GeV are shown in Fig.~\ref{predictionP}.
The resulting picture disfavours both very small and very large values of $r_U$ which, in turn, 
means that one expects {the electron fraction to be neither be small nor too large}
with respect to the positron fraction.  

As the experimental data on the positron fraction should hopefully soon become more accurate \cite{AMS02},
the method outlined above will allow to extract crucial information about the charge asymmetry
of the unknown component of the electrons and positrons in CRs.

\section{Conclusions}
 
The model independent analysis of the combined PAMELA and FERMI-LAT data proposed in ref. \cite{Frandsen:2010mr} 
shows that current data still allow for approximately equal contributions of the electrons 
and positrons from the unknown components of the associated CRs, but disfavor electron to positron 
fractions much smaller than $1/2$ or larger than $4$. 
Hence, the typical oversimplifying model assumption of charge symmetry ($r_U=1$) used so far, 
actually constitutes just a small portion of the allowed models still left unconstrained by the present data.

Interestingly enough, we emphasize that the planned improvements in the experimental sensitivity 
to the positron fraction, as well as further data at energies higher than the one explored so far by PAMELA,
could reveal whether the unknown component of the CR electrons and positrons is charge asymmetric or not.
If it turned out to be charge asymmetric, an explanation in terms of new physics beyond the Standard Model 
would clearly become more compelling with respect to an astrophysical one.
 
\ack We thank F.Palmonari and R.Sparvoli for useful comments and suggestions.

\end{document}